\documentstyle[preprint,pre,aps]{revtex}
\tighten
\begin{document}
\draft
\preprint{\today}
\title{Energy spectra of quasiperiodic systems via information entropy}
\author{Enrique Maci\'{a}$^*$ and Francisco Dom\'{\i}nguez-Adame}
\address{Departamento de F\'{\i}sica de Materiales,
Facultad de F\'{\i}sicas, Universidad Complutense,
E-28040 Madrid, Spain}

\author{Angel S\'{a}nchez}
\address{Escuela Polit\'{e}cnica Superior, Universidad Carlos III,
c.\ Butarque 15, E-28911 Legan\'{e}s, Madrid, Spain}

\maketitle

\begin{abstract}
We study the relationship between the electronic spectrum structure and
the configurational order of one-dimensional quasiperiodic systems.  We
take the Fibonacci case as an specific example, but the ideas outlined
here may be useful to accurately describe the energy spectra of general
quasiperiodic systems of technological interest.  Our main result
concerns the {\em minimization} of the information entropy as a
characteristic feature associated to quasiperiodic arrangements.  This
feature is shown to be related to the ability of quasiperiodic systems
to encode more information, in the Shannon sense, than periodic ones.
In the conclusion we comment on interesting implications of these
results on further developments on the issue of quasiperiodic order.
\end{abstract}

\narrowtext

The notion of {\em quasiperiodic order} (QPO) has been given a very
different status in physics, depending on the perspective adopted by
authors working on different fields.  From a theoretical point of view,
the necessity to achieve a good understanding of the transport
properties of disordered systems has led a number of researchers to
address the issue of quasiperiodicity as somewhat intermediate between
periodic order and purely random disorder \cite{Ko83}.  From this
perspective the notion of QPO assumes a subsidiary role as a mere way to
describe the {\em conceptual} transition from periodic order to
randomness.  On the other side, from a practical viewpoint, it has been
progressively realized, starting from the pioneering works by Merlin et
al. \cite{Merlin} and Todd et al. \cite{Todd} on quasiperiodic
superlatices, that electronic devices with this particular kind of
structure offer interesting possibilities for technological
applications.  In this case the study of those characteristic features
directly related to the underlying QPO becomes interesting by its own
right.

The electronic energy spectra of quasiperiodic systems has been
extensively studied in the framework of real-space renormalization group
techniques \cite{Niu,Liu,Oh}, where the original system is decoupled
into a certain number of minor subsystems according to a given, so
called, {\em blocking scheme} (BS).  To date, two BSs have been proved
to be remarkably useful in describing on-site and transfer models within
the tight-binding approximation \cite{Liu}.  Less atention has been
paid, however, to more general {\em mixed} models, in which both
diagonal and off-diagonal terms are taken into account in the system
Hamiltonian.  Such models are, of course, more appropriate in order to
describe realistic systems, where interactions between their different
constituents depend on their specific chemical nature.

In this Rapid Communication we report on two general results.  On one
side, we show that the BS originally proposed to study on-site models
can be extended to describe mixed models, hence indicating its
applicability to a wide variety of quasiperiodic systems.  On the other
side, we show that this fact is related to the behavior of the
information entropy function associated to these systems.  In spite of
our main conclusions holding for general one-dimensional systems
exhibiting QPO, in this work we focus on Fibonacci arrangements for the
sake of illustration.  The reasons for this particular choice are
twofold.  In the first place, energy spectra corresponding to Fibonacci
systems have been experimentally probed in a variety of situations,
confirming that Fibonacci arrangements exhibit spectra with a hierarchy
of splitting minibands displaying self-similar patterns
\cite{Laruelle,Nakayama,Kono,Tuet}, even when relativistic effects are
taken into account \cite{FDA}.  In addition we have shown recently that
this fractal structure of the energy spectrum has relevant consequences
on the dc conductance of the system \cite{us}.  In the second place, it
is currently well established the widespread appearance of ordering
patterns and structures based on the Fibonacci sequence in many
biological systems \cite{Dou,Levi}.  Therefore, the basic ideas
inspiring our work can be applied to a broad kind of interesting
systems.

Let us start by introducing our model Hamiltonian.  We shall consider a
system describing a binary substitutional alloy in which the constituent
atoms are arranged according to the Fibonacci sequence.  In general, a
Fibonacci chain of order $N$ is generated from two basic units A and B
by successive applications of the substitution rule A~$\rightarrow$~AB
and B~$\rightarrow$~A yielding a sequence of the form ABAABABA \ldots\
This sequence comprises $F_{N-1}$ elements A and $F_{N-2}$ elements B,
$F_l$ being the {\em l\/}th Fibonacci number given by the recurrent law
$F_l=F_{l-1}+F_{l-2}$ with the initial values $F_0=F_1=1$.  In an actual
alloy the hopping integrals describing the interaction between
nearest-neighbor atoms would take different values depending upon
the chemical nature of atomic species.  In order to take into account
this fact, we propose a general {\em mixed} model in the one-electron
approximation.  The interaction of an electron with the host atoms is
described by means of delta function potentials.  This is not a serious
limitation since the delta function is a good approximation to more
realistic short-ranged potentials \cite{delta}.  Therefore, we consider
the following Schr\"odinger equation in units such that $\hbar = m=~1$
\begin{equation}
\left[ -\, {1\over 2}{d^2\over dx^2} -
\sum_{n}\lambda_{n} \delta (x-nd) \right] \psi(x) = E \psi(x),
\label{Sch}
\end{equation}
where we are dealing with equally spaced atomic positions, $d$ being the
\ nearest-neighbor atomic distance.  We allow the potential strength
$\lambda_n$ to take on two values, $\lambda_A$ and $\lambda_B$, arranged
according to the Fibonacci sequence.  Hereafter we restrict ourselves to
attractive potentials ($\lambda_n>0$) and take $\lambda_A=1$ without
loss of generality.  Expressing the electron wave function as a linear
combination of atomic orbitals $\psi(x)=\sum_n C_n \phi_n(x-nd)$, where
$\phi_n(x-nd)=\sqrt{\lambda_n}\exp( -\lambda_n|x-nd|)$ is the normalized
eigenfunction of a delta function placed at $x=nd$, and neglecting the
overlap involving three different centres, we obtain the following
tight-binding equation \cite{us}
\begin{equation}
(E-\epsilon_n)C_n=t_{n,n+1}C_{n+1}+t_{n,n-1}C_{n-1}, \label{tight}
\end{equation}
where the on-site energies and the hopping integrals are given by
\begin{eqnarray}
\epsilon_n&=&-\,{1\over 2}\lambda_n^2, \label{esite} \\
t_{n,n\pm1}&=&-\sqrt{\lambda_n^3\lambda_{n\pm 1}}
\exp (-\,\lambda_{n\pm 1}d).  \label{hopping}
\end{eqnarray}

A complete description of the resulting electronic spectrum can be found
elsewhere \cite{us}.  Here we shall merely give a brief summary of those
results relevant to our purposes.  In our numerical simulations we have
studied in detail different realizations by varying the chain length,
$N$, the interatomic distance, $d$, and the ratio $\alpha \equiv
\lambda_B/\lambda_A$ which accounts for the chemical diversity of the
alloy.  In {\em all} cases considered we have observed a well
differentiated tetrafurcation pattern of the energy spectrum,
characterized by the presence of four main subbands separated by
well-defined gaps.  Inside each main subband the fragmentation scheme
follows a trifurcation pattern in which each subband further trifurcates
obeying a hierarchy of splitting from one to three subsubbands.  In
Fig.\ ~1 we plot a typical resulting integrated density of states (IDOS)
per unit length.  Its stair-case structure clearly shows the
tetrafurcation of the spectrum as well as its self-similarity.  At this
point we shall exploit the capability of the IDOS to provide a link
between the occupation of the various minibands appearing in the
spectrum and the underlying atomic arrangement.  In fact, since each
atom in the chain contributes with one electronic state to the energy
spectrum, the fraction of states is directly related to the heights of
the characteristic steps appearing in the IDOS.  In this way, we have
measured the fraction of states appearing in the main subbands, $q_i$,
for a wide range of both $\alpha$ and $d$ parameters.  The measured
heights yield the same values for all the model parameters considered.
These values agree within an error less than $0.1\%$ to $q_a=q_c=
\tau^3$ , $q_b= \tau^4$ and $q_d= \tau^2$, where $\tau=(\sqrt{5}-1)/2$
is the golden mean.  These values completely agree with those obtained
by Liu and Sritrakool for the IDOS corresponding to on-site models,
which were interpreted in terms of a BS associated to the existence of
isolated A and B atomic blocks and AA diatomic blocks \cite{Liu}.  In
that models, the hierachical splitting of the spectrum is determined by
short-range effects, and its overall structure is due to the resonant
coupling among states with nearly the same energy.  As a result, the
number of energy levels appearing at the first stage of the
renormalization process determines the number of main subbands in the
corresponding spectrum.  Keeping this in mind, the convenience of a
renormalization approach in our mixed model is justified by the
following fact: we have observed that both the position and widths of
the main subbands of the spectrum converge rapidly to stable values with
increasing chain length.  This behavior, which is independent of the
chemical diversity of the alloy, has been referred to as {\em asymptotic
stability} of the spectrum \cite{us} and it seems to be a quite general
property of both quasiperiodic \cite{Baake} and homogeneously disordered
alloys \cite{Cordelli}.  The asymptotic stability hence suggests that
the number of main subbands in the spectrum is determined by short-range
quantum effects, as it is required.  Therefore, according to the
renormalization group ideas just exposed, our obtained IDOS patterns
indicate that we must block the original Fibonacci chain into a secuence
of isolated A and B atomic blocks and diatomic AA blocks in order to
obtain the purported step heights.  Thus, we arrive to our first main
result, namely, that a blocking scheme, originally proposed to account
for the energy spectrum of a very particular kind of system, can be
extended to properly describe more complex and realistic quasiperiodic
systems.

Now, this interesting result deserves some consideration, since it
indicates that the electronic spectrum of these more general systems can
still be accounted for by decimating the original chain in terms of
isolated A and B atomic blocks and AA diatomic blocks, even if one
allows for a significative AB interaction, as we do in our model.  This
striking result strongly suggests that we are facing a property
associated to the QPO, rather than a feature attributable to the
particular choice of the parameters describing the considered model.  In
this regard, we feel it has not been previously stressed in the
literature the fact that, for any arbitrary chain, quasiperiodic or not,
there exist a manifold of possible BSs which may be considered in
principle.  Hence, the question arises as to whether a particular scheme
will be more appropriate than other in order to describe the energy
spectrum splitting pattern.  In the remaining we shall show that, for
systems exhibiting a QPO based on the application of a substitution
rule, the more appropriate BSs are those which {\em minimize} their
associated information entropy function, defined as follows.

Let us consider a general binary system composed of two different
species, A and B. Although we shall consider these species to be
completely general in nature, we shall refer to them as 'atoms' for
convenience henceforth.  According with the above presented numerical
results, it does not seem to be necessary to consider blocks containing
more than two atoms in the first stage of the renormalization process,
within the weak-bound approximation.  Then, the most complete BS to be
considered should involve isolated A and B atomic blocks and AA, AB, BA
and BB diatomic blocks.  As stated before, there are many ways in which
we can decouple the original chain in a series of atomic and diatomic
blocks, each one producing a particular BS. Therefore, we can assign to
each particular BS a probability set of the form $\{p_A, p_B, p_{AA},
p_{AB}, p_{BA}, p_{BB}\}$, where $p_j$ is the probability of finding a
given atomic or diatomic block along the blocked chain.  In the
thermodynamic limit these probabilities are obtained as
$p_j=\lim\,(n_j/N)$, where $n_j$ is the number of $j$-type blocks and
$N$ the total number of atoms in the chain.  In this way, we can
consider the set of possible BSs as an {\em statistical ensemble} in the
usual sense and associate to it an information entropy function given by
\begin{equation}
S = -k\sum_j p_j \ln p_j,
\label{infentrop}
\end{equation}
where $k$ is an appropiate constant.  It then results that each possible
BS can be properly characterized by its information entropy.  Now, it is
clear that the entropy function (\ref{infentrop}) will also depend on
the kind of order exhibited by the alloy we are considering.  Actually,
it becomes that such dependence is very especial for systems exhibiting
QPO. Let us consider, as an example, the case of Fibonacci QPO. In order
that the probability set $\{p_j \}$ can describe a Fibonacci sequence we
must impose certain {\em constrains} onto the possible values of the
different probabilities appearing in it, namely, these probabilities
must satisfy the well-known Fibonacci limits, $\lim\, (n_B/n_A)= \tau$,
and $\lim\, (n_A/n_{AA})=\tau$, along with the normalization condition
$p_A+p_B+2(p_{AA}+p_{AB})=1$, which implies that $p_{AB}=p_{BA}$ and
$p_{BB}=0$ in a Fibonacci chain.  In doing so, we get the following
relationships
\begin{eqnarray}
p_A &=& \tau^3 (\tau - p_{AB}), \nonumber \\ p_B &=&
\tau^2 - p_{AB}, \nonumber \\ p_{AA} &=& \tau^2 (\tau - p_{AB}).
\label{seto}
\end{eqnarray}
where, for convenience, we have left $p_{AB}$ as a free parameter.  By
inspection of expressions (\ref{seto}) we see that the condition $p_j
\geq 0$ implies $0 \leq p_{AB} \leq \tau^2$ and that $p_A$ and $p_B$
{\em cannot} be simultaneously zero.  As a consequence, the original
Fibonacci chain cannot be decoupled into any series of diatomic AB and
AA blocks alone.  Thus, we are led to the conclusion that certain BSs
are prevented by the QPO exhibited by the system.  Alternatively, it
seems reasonable then that certain BSs might be favoured by the QPO. In
order to prove this point we evaluate the information entropy function
associated to the Fibonacci chain making use of the relations
(\ref{seto}).  In Fig.\ ~2 we present its dependence on the probability
$p_{AB}$.  The entropy curve exhibits a characteristic maximum and two
minima at the extreme points of its domain.  The minimum at $p_{AB}=0$
corresponds to the level populations observed in our IDOS and it is
related to the BS proposed by Liu and Sritrakool \cite{Liu}.  The
minimum at $p_{AB}= \tau^2$ is associated to the consideration of three
kinds of blocks as well, namely, A isolated atomic blocks and AA and AB
diatomic blocks.  The corresponding level population can be obtained
from (\ref{seto}) making use of the relationship
$\tau^n+\tau^{n-1}=\tau^{n-2}$ \ ($n>2$), and is given by the sequence
$\tau^5: \tau^2: \tau^6 : \tau^5: \tau^2$.  No evidence has been
observed in our numerical simulations for this pentafurcation pattern of
the IDOS.  The reason for this absence becomes clear from inspection of
Fig.\ 3, where we show the behavior of the diatomic block energies,
$E_{AA}$ and $E_{AB}$, as a function of the atomic distance for a fixed
value of the chemical diversity, $\alpha$.  In fact, we see that energy
levels associated to the blocks AA are always below the levels
corresponding to the blocks AB, hence indicating that the AB interaction
is not strong enough to force all B atomic levels to form AB diatomic
blocks in our model.

In this way we arrive to our second main result, namely, that the most
favorable BSs for a mixed Fibonacci system are those which {\em
minimize} their associated information entropy.  Although this result
has been obtained for a particular kind of QPO, we conjecture that the
principle of minimum information entropy as a criteria to properly
choose the most appropriate BS, may hold for a broader class of
quasiperiodic systems as well.  The reasons sustaining our assumption
rely on the following facts.  In the first place, quasiperiodic systems
do encode {\em more information}, in the sense of Shannon, than periodic
or random ones, since the information content of a periodic chain is
independent of length, whereas that of a quasiperiodic system {\em
increases} with length.  As a consequence, any quasiperiodic system will
content more information than a periodic one if it is sufficiently long
, supporting our view that systems exhibiting QPO should more
appropriately be described in terms of minima of the associated
information entropy.  In the second place, from the IDOS shown in Fig.\
1 we can see that high energy levels are more populated than low energy
ones at zero temperature.  This remarkable feature also appears in the
IDOS describing other aperiodic systems generated by the application of
substitution rules, like Thue-Morse, period-doubling or Rudin-Shapiro
\cite{Luck}.  Therefore, it is tempting to say that QPO describes {\em
far from thermodynamical equilibrium} systems.  The obtention of
one-dimensional quasicrystals exhibiting short Fibonacci stacking
microstructure by means of nonequilibrium methods, such as rapid
quenching from the melt \cite{He} as well as the growth of Fibonacci
superlattice heterostructures by molecular beam epitaxy \cite{Merlin},
supports this point of view.  Now, it is accepted that the information
entropy equals the thermodynamic entropy for equilibrium systems.  There
is no claim, however, that the information entropy represents the
correct expression for the thermodynamic entropy of a system that is not
in equilibrium (or that it does not) \cite{Robertson}.  In this regard
quasiperiodic systems could provide a suitable example of
non-equilibrium systems for which the information entropy is quite
different from its (vanishing) configurational entropy.

In summary, we have obtained valuable results concerning the
relationship between the electronic spectrum structure and the
information entropy content of a general Fibonacci system describing a
binary alloy.  Our results strongly support the view that QPO describes
highly ordered systems able to encode more information than cristalline
ones.  By introducing a novel, general approach to the study of
quasiperiodic systems we provide a link between the atomic arrangement
and the electronic structure displayed by the IDOS, indicating that the
QPO describes systems far from equilibrium.  Therefore , the
maximization entropy formalism, which has proved to be a good strategy
of optimal prediction for both periodic and random systems, does not
seem to work well for quasiperiodic ones.  Finally, the general
treatment introduced in this work can be extended, in a straightforward
manner, in order to describe energy spectra of quasiperiodic systems
other than the Fibonacci one.  Work in this sense, regarding both
electron and phonon energy spectra of systems displaying QPO, is
currently in progress, and we expect to report on it elsewhere.

This work is partially supported by Universidad Complutense de Madrid
under project $4811$. We thank M.\ V.\ Hern\'{a}ndez for a critical
reading of the manuscript.

\begin{figure}
\caption{Plot of the IDOS versus energy with model parameters given by
$\alpha=0.75$, $d=1.5$ and $N=987$.  The main subbands are labelled by
{\em a,b,c} and {\em d} along with their respective fraction of states.
Note that the level population exhibits characteristic features of a
non-equilibrium system.}
\label{fig1}
\end{figure}
\begin{figure}
\caption{Plot of the information entropy curve versus the probability
$p_{AB}$.  The appropriate blocking schemes accomodate to the minima of
entropy function.  See more details in the main text.}
\label{fig2}
\end{figure}
\begin{figure}
\caption{Plot showing the behavior of diatomic block energies
$E_{AA}^{\pm}$ (solid lines) and $E_{AB}^{\pm}$ (dashed lines) with
interatomic distance $d$ for $\alpha=0.75$. Energies corresponding
to isolated A and B atoms are also plotted for comparison (dotted
lines).}
\label{fig3}
\end{figure}

\end{document}